\documentclass[prd,a4paper,twocolumn,superscriptaddress,nofootinbib,showpacs]{revtex4}

\def\npb#1#2#3{{\it Nucl. Phys.} {\bf B#1} (#2) #3 }
\def\plb#1#2#3{{\it Phys. Lett.} {\bf B#1} (#2) #3 }
\def\prd#1#2#3{{\it Phys. Rev. } {\bf D#1} (#2) #3 }
\def\prl#1#2#3{{\it Phys. Rev. Lett.} {\bf #1} (#2) #3 }
\def\mpla#1#2#3{{\it Mod. Phys. Lett.} {\bf A#1} (#2) #3 }

\def\bb#1{{\tt hep-th/#1}}
\def\grqc#1{{\tt gr-qc/#1}}
\def\heph#1{{\tt hep-ph/#1}}

\def\rmp#1#2#3{{\it Rev. Mod. Phys.} {\bf #1} (#2) #3 }

\def\cqg#1#2#3{{\it Class. Quantum Grav. } {\bf #1} (#2) #3 }

\def\jhep#1#2#3{{\it J. High Energy Phys.} {\bf #1} (#2) #3 }

\def\aph#1{{\tt astro-ph/#1}}
\def\nc#1#2#3{{\it Nuovo Cimento } {\bf #1} (#2) #3 }

 \def\CC{{\cal C}}  \def\CG{{\cal G}}
\def\CL{{\cal L}}   
 \def\CR{{\cal R}}  \def\CT{{\cal T}}
\def\CM{{\cal M}}

\def\dj{\hbox{d\kern-0.347em \vrule width 0.3em height 1.252ex depth
-1.21ex \kern 0.051em}}

\def\half{{1\over 2}\,}

\def\ee{{\rm e}\,}

\begin{document}

\title[Curved dilatonic brane worlds]{Curved dilatonic brane worlds}

\author{A. Feinstein}
\email{wtpfexxa@lg.ehu.es}
\affiliation{Dpto. de F\'{\i}sica Te\'orica, Universidad del Pa\'{\i}s
Vasco, Apdo. 644, E-48080 Bilbao, Spain \\}
\author{K.E. Kunze}
\email{kunze@amorgos.unige.ch}
\affiliation{D\'epartement de Physique Th\'eorique, 
Universit\'e de Gen\`eve, 24 Quai Ernest Ansermet, CH-1211 Gen\`eve 4, Switzerland \\}
\author{M.A. V\'azquez-Mozo}
\email{vazquez@nbi.dk}
\affiliation{Niels Bohr Institute, Blegdamsvej 17, DK-2100 Copenhagen \O, Denmark \\}

\begin{abstract}

\vspace*{0.6cm}

We construct a broad family of exact solutions to 
the five-dimensional Einstein equations coupled to a scalar field with an exponential potential. 
Embedding a three-brane in these bulk space-times 
in a particular way we obtain a class of self-tuned curved brane worlds in which 
the vacuum energy on the brane is {\it gravitationally idle}, the four-dimensional geometry
being insensitive to the value of the brane tension. This self-tuning arises from cancellations,
enforced by the junction conditions, between the scalar field potential, the brane vacuum energy and the 
matter on the brane. Finally, we study some physically relevant examples and their dynamics. 

\vspace*{0.5cm}

\noindent
{\footnotesize Preprint numbers: EHU-FT/0103, {\tt hep-th/0105182}}

\end{abstract}

\preprint{EHU-FT/0103}
\preprint{hep-th/0105182}

\maketitle

\section{Introduction}

Brane worlds, the models in which our universe appears as some kind of domain wall or brane
embedded in a higher dimensional space-time, have gained considerable attention recently.
Although the idea goes back to the eighties \cite{rubshap}, its current appeal comes from 
several directions. As shown by Ho\v{r}ava and Witten \cite{hw}, 
the $E_{8} \times E_{8}$ heterotic string theory at strong coupling is described in terms 
of M-theory in an eleven-dimensional space-time with boundaries, where the ten-dimensional 
gauge degrees of freedom live on the ``branes at the end of the world''. This confinement
of the gauge fields to a lower-dimensional submanifold, in contrast to the gravitational field
that can propagate into the bulk, might help to explain the hierarchy between the 
electroweak and the Planck scale in four dimensions \cite{lardim}. With the aim of
solving the hierarchy problem, Randall and Sundrum \cite{rs}
put forward a proposal, inspired by
the AdS/CFT correspondence, where our four-dimensional universe is embedded in a non-factorizable
way into five-dimensional anti-de Sitter (AdS) space-time. Gravity in this scenario is ``trapped'' 
on the four-dimensional brane due to the geometry of the bulk space-time \cite{rs,gt}.

The phenomenological viability of brane world models  is now being extensively discussed in the 
literature \cite{phen}. Besides the possible imprints of these models detectable in high energy 
experiments, cosmology naturally emerges as a very promising arena to study  the possible consequences
of living inside a brane (for an incomplete list of references see \cite{bc,mwbh,cll}). 
One of the important issues in  cosmology is 
to explain how our homogeneous and isotropic universe could have
emerged from ``generic'' initial conditions. A popular mechanism to address this question is
to have a certain period of inflation in the early universe, during
which the possible initial anisotropies and inhomogeneities would be ironed off.
In the context of the brane world the issue of inflation 
\cite{mwbh,cll,hl}, as well as the perturbative
deviations from isotropy \cite{mss,svf} have been recently investigated. 

One of the most interesting consequences of 
considering our universe as a brane living inside a higher-dimensional space time is that
Einstein equations in four dimensions do not form a closed system \cite{mss}. As a
consequence, for a four-dimensional observer it is not sufficient to know the distribution of 
energy-matter in her/his universe to determine its geometry,  the missing element
coming from the geometrical features of the space-time outside the four-dimensional
universe. This ``out of this world'' ingredient to the right-hand side of the Einstein equations
is crucial in analyzing the cosmological dynamics of the universe in four-dimensions. However, 
in many instances in the literature the five-dimensional solution in the bulk associated with a
four-dimensional brane cosmology is not known and its effects on the brane world have 
to be modeled 
using some simplifying  assumptions, or neglected altogether.  

Thus, in order to gauge to what extent the five-dimensional geometry influences the cosmological
dynamics on the brane, it is important to consider exact bulk solutions with various deviations
from homogeneity and isotropy (see, for example, \cite{hlz,fr,csop} for some 
studies in this direction). In this paper we will propose a systematic way of constructing
five-dimensional homogeneous and 
inhomogeneous cosmologies coupled to a  scalar field with an exponential 
potential. We will construct brane cosmologies using these bulk metrics and study the effect
that the bulk dynamics has
on the cosmological evolution of the brane world. 

The cosmological constant problem remains a central issue to be solved in theoretical physics
\cite{wein}.
Brane cosmology provides new approaches that might help in the solution of this long-standing
problem \cite{strau}. One of the proposals recently put forward is a self-tuning mechanism which tunes the
four-dimensional cosmological constant to zero, independently of the value of the cosmological
constant in the bulk \cite{kach,kachh,nem,cm}. 
In this paper we propose rather a new realization of this self-tuning
property in which vacuum energy on the brane cancels despite the ``bare'' four-dimensional
cosmological constant being non-zero. This is due to a non-trivial counterbalance between
the non-vanishing ``bare'' cosmological term and the matter induced on the brane.

The paper is outlined as follows. In the next Section we will briefly describe the dynamics of 
dilaton brane cosmologies. After this, in Section 3, we  present a solution generating 
technique to build five-dimensional scalar cosmologies with exponential potential 
starting with a vacuum solution in five dimensions.
Section 4 will be devoted to the study of brane cosmologies embedded in the class of five-dimensional
metrics obtained here, and how for these particular embeddings there is a self-tuning mechanisms at
work, the four-dimensional geometry being independent of the vacuum energy on the brane and the
value of the scalar field potential. In Section 5 we illustrate our discussion with some physically 
interesting examples, and in Section 6 we summarize our results.

\section{Brane dynamics}

In the spirit of the brane world picture, we  assume that the four-dimensional universe
is described by a domain wall $(M,g$) located at some hypersurface $Y(x^{A})=0$ 
in the five-dimensional bulk space-time $(\CM,\CG)$. The only matter in the bulk
will be a massless scalar field with an exponential potential. Therefore, the action 
governing the dynamics is\footnote{In the following we will
use capital Latin indices for the five dimensional coordinates, whereas the coordinates
on the brane world will be denoted by Greek indices. To avoid the use of superindices to 
indicate the dimension, tensors in five dimensions will be indicated by capital caligraphic
letters and their four-dimensional counterparts will be denoted by the corresponding
roman types.} 
\begin{eqnarray}
S_{\rm 5D} &=& \int d^{5}x\sqrt{-\CG}\left[{1\over 2\kappa_{5}^{2}}
\CR-\half\partial_{A}\phi\partial^{A}\phi
-\Lambda e^{-{2\over 3}k\phi}\right] \nonumber \\
&+&\int_{Y=0} d^{4}x\sqrt{-g}L_{\rm brane} \, ,
\label{act}
\end{eqnarray}
where $k$, and $\Lambda$ are constants and 
$$
L_{\rm brane}=-\lambda(\phi)+{1\over \kappa_{5}^2}K^{\pm}+e^{4b\phi}\,L(e^{2b\phi}g_{\mu\nu},\ldots)_{\rm matter} \, .
$$
Here $\lambda(\phi)$ is a $\phi$-dependent vacuum energy 
(i.e. tension) on the brane, $K^{\pm}$ is the extrinsic curvature on either side of the brane 
and $L_{\rm matter}$ is the Lagrangian of the matter degrees 
of freedom confined to the brane world 
minimally coupled to the metric $e^{2b\phi}g_{\mu\nu}$, $b\in{\bf R}$ (cf. \cite{mw}). 
In the following we  use units in which $\kappa_{5}=1$.
The induced metric on the brane is the projection of the five-dimensional metric onto the
brane-world. If we denote by $n^{A}$ the unit space-like vector normal to the brane, the 
four-dimensional metric will be given by
$$
g_{AB}=\CG_{AB}-n_{A}n_{B}\,.
$$
Because of the embedding 
there will be two sources of curvature for the four-dimensional universe. One will be 
the intrinsic curvature induced by the ambient space-time and it will be given by the
projection of the five-dimensional Riemann tensor onto the brane. The second one is due to the 
embedding itself and it is governed by the extrinsic curvature $K_{\mu\nu}=g_{\mu}^{C}g_{\nu}^{D}
\nabla_{C}n_{D}$. Using Gauss-Codazzi equations
\cite{wald} we can write the Einstein equations in four dimensions as \cite{sms,mw,mb} (for
a review see \cite{roy})
\begin{eqnarray}
R_{\mu\nu}&-&\half g_{\mu\nu}R = {2\over 3}\left[\CT_{AB}g_{\mu}^{A}
g_{\nu}^{B} \right.\nonumber \\
&+&\left.\left(\CT_{AB}n^{A}n^{B}-{1\over 4}g^{AB}\CT_{AB}\right)g_{\mu\nu}\right]
+ K\,K_{\mu\nu} \nonumber \\
&-&K^{\,\,\sigma}_{\mu}K_{\nu\sigma}-\half g_{\mu\nu}(K^2-K^{\alpha\beta}K_{\alpha
\beta})-E_{\mu\nu}
\label{ee}
\end{eqnarray}
where we have denoted by $K\equiv K_{\mu}^{\mu}$ the trace of the second fundamental form, 
$E_{\mu\nu}$ is written in terms of the five-dimensional Weyl tensor $\CC^{A}_{\,\,\,\,BCD}$ as
$$
E_{\mu\nu}=\CC^{E}_{\,\,\,\,AFB}\,n_{E}n^{F}g^{A}_{\mu}g^{B}_{\nu}
$$
and the five-dimensional energy-momentum tensor in the bulk derived from (\ref{act}) is
\begin{equation}
\CT_{AB}=\partial_{A}\phi\partial_{B}\phi-\CG_{AB}\left(\half \partial_{C}\phi
\partial^{C}\phi+\Lambda e^{-{2\over 3}k\phi}\right)\,.
\label{em}
\end{equation}
It is important to point out that the right-hand side of eq. \ee\ can be evaluated on any side of the brane,
the Einstein tensor on the brane being uniquely defined.  In the 
${\bf Z}_{2}$-symmetric case to be studied below both the extrinsic curvature and the derivatives of the
dilaton field on the two sides of the brane differ just by a sign, so every term in the right hand side of \ee\ is
well defined on the brane. If the reflection symmetry is relaxed, on the other hand, the uniqueness of the 
Einstein tensor is ensured in a nontrivial way \cite{bcmu}.

In addition to this, the scalar field will satisfy the wave equation
\begin{equation}
\nabla^2\phi+{2k\over 3}\Lambda e^{-{2k\over 3}\phi}={\sqrt{-g}\over \sqrt{-\CG}}
\left[\lambda'(\phi)-b\,g^{\mu\nu}\tau_{\mu\nu}
\right]\,\delta(Y)
\label{dilat}
\end{equation}
where $\tau_{\mu\nu}$ is the energy momentum tensor of the matter action, as derived from 
$e^{4b\phi}L_{\rm matter}$, and
the prime denotes differentiation with respect to $\phi$.
Upon projection we obtain the four-dimensional equation for the dilaton
\begin{eqnarray}
D_{\mu}D^{\mu}\phi &-& a_{C}\partial^{C}\phi+K\CL_{n}\phi+\CL_{n}^2\phi+{2k\over 3}\Lambda\,
e^{-{2\over 3}k\phi} \nonumber \\
&=&{\sqrt{-g}\over \sqrt{-\CG}}\left[\lambda'(\phi)-b\,g^{\mu\nu}\tau_{\mu\nu}
\right]\,\delta(Y)
\end{eqnarray}
with $D_{\mu}$ the covariant derivative with respect to the induced metric, 
$a^{C}=n^{B}\nabla_{B}n^{C}$ and $\CL_{n}\phi=n^{A}\nabla_{A}\phi$ the Lie derivative in the direction 
$n^{A}$.

As usual, we will take coordinates $(\chi,x^{\mu})$ in such a way that the braneworld lies on the 
hypersurface defined by $\chi=0$. For later convenience we will consider that the 
five-dimensional metric takes the form
$$
ds^{2}_{\rm 5D} = N(x,\chi)^{2}d\chi^2+g_{\mu\nu}(x,\chi)\,dx^{\mu}dx^{\nu}
$$
where the ``shift'' function $N(x,\chi)$ depends on all five-dimensional coordinates. Therefore
$n\equiv N(x,\chi)^{-1}\partial_{\chi}$ and as a consequence 
\begin{eqnarray}
K_{\mu\nu}&=&{1\over 2N(x,\chi)}\partial_{\chi}g_{\mu\nu}(x,\chi) \nonumber \\
a_{\mu}&=&
-{1\over N(x,\chi)}\partial_{\mu}N(x,\chi),\hskip 1cm a_{\chi}=0 \nonumber 
\end{eqnarray}
and $E_{\mu\nu}=\CC^{\chi}_{\,\,\,\,\mu\chi\nu}$. 

The discontinuity of the derivatives of the metric across the brane due to the energy-momentum 
localized on the hypersurface $Y(x^{A})=0$ is given by the Israel junction conditions 
\cite{is,sms}. They relate the jump in the first derivative of the metric at $\chi=0$ with the 
total energy-momentum tensor on the brane,namely
\begin{equation}
\left[K_{\mu\nu}\right]=-\left(S_{\mu\nu}-{1 \over 3}g_{\mu\nu}S\right)
\label{sk}
\end{equation}
where we have used the usual notation $[A]\equiv A^{+}-A^{-}$, and
 $S_{\mu\nu}$ is the total brane energy-momentum tensor 
\begin{equation}
S_{\mu\nu}=-{2\over \sqrt{-g}}{\delta\over \delta g^{\mu\nu}}(\sqrt{-g}\,L_{\rm brane})
\equiv -\lambda(\phi)g_{\mu\nu}+\tau_{\mu\nu}.
\label{s}
\end{equation}
In a similar fashion we can find the jump in the derivative of the scalar field  
by integrating (\ref{dilat}) across $\chi=0$, namely
\begin{equation}
\left[\,\partial_{\chi}\phi\,\right]
= N(x,0)\left[\lambda'(\phi)-
b\,g^{\mu\nu}\tau_{\mu\nu}\right].
\label{matdil}
\end{equation}
The Einstein equations in four dimensions can be now written from
(\ref{ee}) as
\begin{eqnarray}
\hspace*{-0.2cm}
R_{\mu\nu}&-&\half g_{\mu\nu}R ={2\over 3}\left[\partial_{\mu}\phi\partial_{\nu}\phi+{5\over 8}
(\partial\phi)^2 g_{\mu\nu}\right] \nonumber \\
&+&{(\partial_{\chi}\phi)^2\over 4N(x,0)^2}g_{\mu\nu}-\half \Lambda\,
e^{-{2\over 3}k\phi}\,g_{\mu\nu} 
+ K\,K_{\mu\nu}\nonumber \\
&-&K_{\mu}^{\,\,\,\sigma}K_{\nu\sigma}
-\half\,g_{\mu\nu}\left(K^2-K^{\alpha\beta}
K_{\alpha\beta}\right)-E_{\mu\nu}.
\label{mee}
\end{eqnarray}
As we discussed above, the only discontinuities on the right-hand side of (\ref{mee}) are contained in
the extrinsic curvature $K_{\mu\nu}$, the derivative of the dilaton field ``normal'' to the brane
$\partial_{\chi}\phi$ and, eventually, the potential. These terms have to
be evaluated at any side of the
brane, the sum of them being independent of the side chosen. All other terms, involving the 
scalar field and their ``tangent'' derivatives are continuous and can be evaluated individually without 
ambiguity on the hypersurface $\chi=0$.

If we assume that our brane world is at the fixed point of a ${\bf Z}_{2}$ orbifold, as it is 
the case in the Ho\v{r}ava-Witten scenario, $K_{\mu\nu}^{+}=-K_{\mu\nu}^{-}$ and the Israel 
junction condition (\ref{sk}) completely determines the extrinsic curvature in terms of the energy-momentum 
tensor on the brane. In the same way $(\partial_{\chi}\phi)^2$ can be read
off eq. (\ref{matdil}).  Substitution 
into (\ref{mee}) leads then to (cf. \cite{mw})
\begin{eqnarray}
R_{\mu\nu}&-&\half R\,g_{\mu\nu}={2\over 3}\left[\partial_{\mu}\phi\partial_{\nu}\phi-{5\over 8}g_{\mu\nu}
(\partial\phi)^2\right]\nonumber \\
&+&{1\over 6}\lambda(\phi)\tau_{\mu\nu}- \Lambda_{4}g_{\mu\nu} -{1\over 16}[2\lambda'(\phi)
-b\,\tau]\,b\tau\,g_{\mu\nu}\nonumber \\
&+&\pi_{\mu\nu}-E_{\mu\nu},
\label{eef}
\end{eqnarray}
where $\tau\equiv g^{\alpha\beta}\tau_{\alpha\beta}$ and 
the four-dimensional cosmological constant and the tensor $\pi_{\mu\nu}$ are given respectively
by
\begin{eqnarray}
\Lambda_{4}&=& \half\left[\Lambda e^{-{2k\over 3}\phi}+{1\over 6}\lambda(\phi)^2-{1\over 8}\lambda'(\phi)^2\right]
\label{cc} \\
\pi_{\mu\nu} &=& {1\over 12}\tau\,\tau_{\mu\nu}+{1\over 8}g_{\mu\nu}\,\tau_{\alpha\beta}\tau^{\alpha\beta}
-{1\over 4}\tau_{\mu\alpha}\tau^{\alpha}_{\,\,\nu} - {1\over 24}\tau^{2}\,g_{\mu\nu}\,.
\nonumber 
\end{eqnarray}

Looking at the Einstein equations on the brane, eq. (\ref{eef}), we find that in general the dynamics of the
four-dimensional universe is not uniquely determined by the distribution of energy inside the 
universe as encoded in $\tau_{\mu\nu}$ \cite{mss,mw}, as it is the case with ``ordinary'' Einstein equations. 
Indeed, the only ingredient on the right-hand side of (\ref{eef}) which cannot be related with the matter content of the 
four-dimensional universe (i.e. either four-dimensional matter or the scalar field $\phi$) is the tensor $E_{\mu\nu}$
which is determined by the five-dimensional Weyl tensor. In order to study the influence of the five-dimensional
geometry on the cosmological evolution of the brane world via 
$E_{\mu\nu}$ it is necessary to consider not only 
the four-dimensional metric on the brane but also the higher dimensional ambient geometry. Thus, 
to address this problem, we will proceed to construct explicitly five-dimensional solutions to the Einstein equations 
in which our brane worlds will be embedded.

\section{Scalar field cosmologies in five dimensions with an exponential potential}

Since we will be interested in studying the effects of the bulk on 
the four-dimensional brane world our starting point will be the five-dimensional 
geometry in which the brane is  embedded. In particular, we want to 
consider solutions to the equations of motion derived from the bulk terms in eq. (\ref{act}), i.e. geometries
coupled to a massless scalar field with a Liouville potential.

Such cosmologies can be constructed using a higher dimensional generalization of the theorem 
presented in ref. \cite{fonarev}. Related models were studied
by Lidsey \cite{jim}
in the context of heterotic M-theory; some particular examples of constant 
curvature dilatonic branes were also discussed in \cite{ajs}.
 Let us consider a {\it vacuum} solution to the Einstein equations
in five dimensions of the form
\begin{equation}
ds^2_{\rm (vac,5D)} = \epsilon\,e^{4\,Q(x)}d\chi^2+e^{-2\,Q(x)}\,h_{\mu\nu}(x)\,dx^{\mu}dx^{\nu}
\label{sp}
\end{equation}
where $\epsilon=\pm 1$ depending on whether $\chi$ is a spatial direction or the time coordinate and 
the metric functions $Q(x)$ and $h_{\mu\nu}(x)$ are independent of $\chi$-coordinate. The new metric
\begin{eqnarray}
ds^{2}_{\rm 5D} &=& \epsilon\,e^{{4k\over \sqrt{k^2+6}}Q(x)+a_{1}\xi\,\chi} \,d\chi^2\nonumber \\
&+& e^{-{2k\over \sqrt{k^2+6}} Q(x)+
a_{2}\xi\,\chi}\,h_{\mu\nu}(x)\,
dx^{\mu}dx^{\nu} 
\label{lefd}
\end{eqnarray}
and the scalar field
\begin{equation}
\phi(x,\chi)={6\over \sqrt{k^2+6}}\,Q(x)+\left\{
\matrix{{3k\xi\over k^2-3}\chi & & k^2\neq 3 \cr & & \cr
\pm \sqrt{3}\xi\,\chi & & k^2=3 }
\right.
\label{dil}
\end{equation}
solve Einstein equations coupled to a massless scalar field with potential $V(\phi)=\Lambda\, e^{-{2\over 3}k\phi}$
where
$$
\Lambda = \left\{ 
\matrix{-{9\over 2}{12-k^2\over (k^2-3)^2}\xi^{2} & k^2\neq 3 \cr
 & \cr
-{9\over 2}\xi^2 & k^2=3 }
\right.
$$
and
\begin{eqnarray}
a_{1}\,&=&{k^2\over 3}a_{2}\,\equiv\,{2k^2\over k^2-3} \hskip 1.2cm {\rm if} \,\,\,k^2\neq 3\, , \cr
a_{1}\,&=&a_{2}\,\equiv\, 2 \hskip 2.5cm   {\rm if} \,\,\, k^2=3\, .
\label{gam}
\end{eqnarray}
Finally, $\xi$ fixes the scale of the cosmological constant $\Lambda$.

Though some algebra is involved, the result can be easily proven
 following the same steps as in the four-dimensional case \cite{fonarev},  
we therefore  leave this to the reader. Using the  theorem we can construct five-dimensional dilaton 
gravity solutions with a cosmological
constant (negative when $k^{2}<12$). At a glance, there are several interesting values
of $k$ for which the metric (\ref{lefd}) admits different physical interpretations. The first
one is $k^2=12$ when the potential for the scalar field vanishes. The second, and more interesting one, is
$k=0$ at which the coupling between the dilaton field and the cosmological constant is zero and we are left
with a negative cosmological constant $\Lambda=-6\xi^2$. In this case the geometries are characterized by 
the line element
\begin{equation}
ds^2_{k=0}=\epsilon\,d\chi^2+e^{-2\xi \chi}\,h_{\mu\nu}(x)\,dx^{\mu}dx^{\nu}
\label{rsun}
\end{equation}
which solve the Einstein equations with a negative cosmological constant and a massless scalar field 
$\phi(x)=\sqrt{6}Q(x)$.  Taking $\epsilon=1$ and $\chi$ as the bulk coordinate, we are 
provided with generalizations of the 
Randall-Sundrum model with a generic four-dimensional metric and a massless scalar field.
If $k^2=18$ and $\epsilon=1$ we obtain solutions to the low-energy field
equations of the $E_{8}\times E_{8}$ heterotic string at strong coupling compactified
on a three-fold Calabi-Yau, with the scalar
field $\phi$ representing the breathing mode of the internal manifold \cite{losw,jim}.

Whenever $\epsilon=1$ it can be easily realized that the vacuum five-dimensional metric (\ref{sp})
 can be thought of as the 
``oxidation'' of a four-dimensional massless scalar field cosmology with a a scalar field 
$\psi(x)_{\rm 4D}=
\sqrt{6}\,Q(x)$ (cf. for example \cite{fvm}). Therefore in order to get (\ref{sp}) one can start
with {\it any} four-dimensional solution to the Einstein-scalar equations. 
If $k=0$ the previous results tell us that {\it any} four-dimensional metric coupled
to a massless scalar field can be embedded into a five-dimensional bulk space-time with negative cosmological
constant by the ansatz (\ref{rsun}) with $\epsilon=1$.
Finally, in the case $Q(x)=0$ the Einstein vacuum equations for (\ref{sp}) imply that $h_{\mu\nu}$ has 
to be a Ricci flat four-dimensional metric. For the particular choice 
$h_{\mu\nu}=\eta_{\mu\nu}$ we recover the family of metrics considered in \cite{kach} 
after the obvious change of coordinates $dx_{5}=\exp\left({1\over 2}a_{1}\xi\chi\right)d\chi$ and
a rescaling of the dilaton field. On the other hand, taking $h_{\mu\nu}$ to be the Schwarzschild
metric in four-dimensions we can construct embeddings of four-dimensional black holes in five dimensions of 
the form
\begin{eqnarray}
ds^2&=&e^{a_{1}\xi \chi}d\chi^2 \nonumber \\
&+&e^{a_{2}\xi\chi}\left[-\left(1-{2M\over r}\right)dt^2
+{dr^2\over 1-{2M\over r}}+r^2 d\Omega_{2}^2\right], \nonumber 
\end{eqnarray}
which, for $k=0$, corresponds to the AdS black string considered by Chamblin, Hawking and Reall \cite{rh}.
When $k\neq 0$ the four-dimensional black hole is embedded into a five-dimensional bulk space-time with
a nontrivial profile for the dilaton field, which on the other hand is constant on the brane.

\section{Brane cosmologies with idle vacuum energy}

In the previous Section we have constructed a generic procedure to obtain five-dimensional
cosmologies coupled to a scalar field with an exponential potential (or a negative cosmological
constant). The final aim is to use these solutions as bulk geometries of four-dimensional brane worlds.
As we have reviewed in Section 2, given the solution in the bulk, the matter/energy content of the brane world
is strongly constrained by the junction conditions for both the metric and the scalar field. 

Among the different possibilities to embed a four-dimensional brane world in the five-dimensional
solution (\ref{lefd}) the simplest ones correspond to take the codimension of the brane along one of the 
four space-like coordinates.

Let us consider the solutions (\ref{lefd}) and (\ref{dil}) with $\epsilon=1$ so that
the brane world lies on the hypersurface
defined by the equation $\chi=0$. This choice leads naturally to warped geometries that 
generalize the Randall-Sundrum construction to include a scalar field with a Liouville potential.

We begin by assuming ${\bf Z}_{2}$-symmetry around the location of
the brane at $\chi=0$. In this case we are led to the following non-factorizable 
geometry in five dimensions
\begin{eqnarray}
ds^{2}_{\rm 5D} &=& e^{{4k\over \sqrt{k^2+6}}Q(x)+a_{1}\,\xi|\chi|} \,d\chi^2\nonumber \\
&+& e^{-{2k\over \sqrt{k^2+6}} Q(x)
+a_{2}\,\xi|\chi|}\,h_{\mu\nu}(x)\,
dx^{\mu}dx^{\nu} 
\label{warp}
\end{eqnarray}
with $a_{1}$, $a_{2}$ given by eq. (\ref{gam}). The dilaton, on the other hand, is given by
\begin{eqnarray}
\phi(x,\chi)=\varphi(x)+\left\{
\matrix{{3k\xi\over k^2-3}\,|\chi| & & k^2\neq 3 \cr & & \cr \pm \sqrt{3}\,\xi\,|\chi| & & k^2=3 }
\right.
\label{dill}
\end{eqnarray}
where $\varphi(x)={6\over \sqrt{k^2+6}}Q(x)$.

In principle, for generic $k\neq 0$, the metric (\ref{warp}) can have a curvature singularity located 
at $\chi=\pm\infty$ (cf. \cite{chsa,kach}), in addition to the possible singularities of the metric $h_{\mu\nu}$. 
Depending on the value of $k$ and on the particular solution considered 
this singularity can be harmless provided it is located at an infinite proper distance from 
the brane 
location. It can be easily seen that this is the case whenever $k^{2}\geq 3$. However, in this
case the four-dimensional Planck scale obtained by integrating out the coordinate $\chi$ in the 
five-dimensional action (\ref{act}) diverges. On the other hand, when $k^2 < 3$ the singularity
is located at finite proper distance from the brane but the four-dimensional 
Planck scale is finite as well. Whenever needed, a second brane can be located at $\chi_{0}>0$, imposing
reflection symmetry around this point, so the bulk coordinate is restricted to the interval 
$[0,\chi_{0}]$ and the singularity is screened.

Since we have completely determined the bulk geometry and assumed reflection symmetry around
$\chi=0$, the junction conditions (\ref{sk}) will determine the energy-momentum tensor of the matter fields
on the brane. Imposing the matching condition on the scalar field (\ref{matdil}) at the brane location for 
the dilaton (\ref{dill}) we find the following relation
\begin{equation}
\lambda'(\varphi)-b\,g^{\mu\nu}\tau_{\mu\nu}=\left\{
\matrix{{6k\xi\over k^2-3}\,e^{-{k\over 3}\varphi} & & k^2\neq 3 \cr
 & & \cr
\pm 2\sqrt{3}\,\xi e^{\mp{1\over \sqrt{3}}\varphi} & & k^2=3 }\right.
\label{rel}
\end{equation}
where we have used the fact that in our case the shift function $N(x,\chi)$ can be written in terms
of the scalar field as
\begin{equation}
N(x,\chi)= e^{{k\over 3}\phi(x,\chi)}
\label{relat}
\end{equation}

Moreover, using (\ref{rel}) and the expression of the extrinsic curvature of the hypersurface $\chi=0$ 
embedded in the metric (\ref{warp}), 
$K_{\mu\nu}=\half a_{2}\,\xi\,e^{-{k\over 3}\varphi}\, g_{\mu\nu}$, 
we can write a differential equation for $\lambda(\varphi)$
\begin{equation}
\lambda'(\varphi)-4b\,\lambda(\varphi)=\left\{\matrix{6\xi{k+12b\over k^2-3}
\,e^{-{k\over 3}\varphi} & & k^2\neq 3 \cr
 & & \cr
2\xi\left(12b\pm\sqrt{3}\right)e^{\mp{1\over \sqrt{3}}\varphi} & & k^2=3}
\right.
\label{diff}
\end{equation}
This equation can be solved to get the functional dependence of
the vacuum energy on the brane with the dilaton field
\begin{equation}
\lambda(\varphi)=-\rho_{0}\, e^{4b\varphi}-\left\{
\matrix{{18\xi\over k^2-3}\,e^{-{k\over 3}\varphi} & & k^2\neq 3 \cr & & \cr
6\,\xi\, e^{\mp {1\over \sqrt{3}}\varphi} & & k^2=3}
\right.
\label{lamb}
\end{equation}
where $\rho_{0}$ is an integration constant. We can now
plug (\ref{lamb}) into eqs. (\ref{s}) and (\ref{sk}) to obtain the form of the energy-momentum tensor $\tau_{\mu\nu}$.
Proceeding in this way we find
\begin{equation}
\tau^{\mu}_{\nu}= -\rho_{0}\, e^{4b\varphi}\, \delta^{\mu}_{\nu}\, .
\label{emt}
\end{equation}
Therefore the matter content of the brane described by $L_{\rm matter}$ has as its effective equation of state
$p=-\rho_{0}$. Notice that the form of the energy-momentum tensor is independent of the value of $k$ 
and therefore valid for {\it all} brane models constructed using the solution generating technique 
of Section 3, and that  the constant energy density and the equation of state do not depend on 
the value of the dilaton coupling $b$. We now evaluate the tensor
$\pi_{\mu\nu}$ which appears on the right-hand side of (\ref{eef}). The result is
$$
\pi_{\mu\nu}=-{1\over 12}\rho_{0}^2\,e^{8b\varphi}\,g_{\mu\nu}.
$$

With these expressions for the tensors appearing on the right-hand side of (\ref{eef}) we can study the 
source of the brane gravitational field. Remarkably, we find that
\begin{equation}
{1\over 6}\lambda(\varphi)\tau_{\mu\nu}-\Lambda_{4}g_{\mu\nu}-{1\over 16}[2\lambda'(\varphi)-b\,\tau]b\,\tau
g_{\mu\nu}+\pi_{\mu\nu}=0
\label{cancel}
\end{equation}
so the ``effective'' energy-momentum tensor in (\ref{eef}) only receives contributions from the dilaton field
and the tidal bulk effects
$$
T_{\mu\nu}^{\rm eff}= {2\over 3}\left[\partial_{\mu}\varphi\partial_{\nu}\varphi-{5\over 8} g_{\mu\nu}
(\partial\varphi)^2\right]-E_{\mu\nu}.
$$

An interesting thing to notice about the cancellation (\ref{cancel}) is that the 
four-dimensional cosmological constant, as defined by eq. (\ref{cc}), is different from zero. 
Only when $\rho_{0}=0$ and $Q(x)=0$ the four-dimensional cosmological constant $\Lambda_{4}$
vanishes exactly independently of the value of the cosmological constant in the bulk.
A particular case is the Randall-Sundrum model \cite{rs}, where
$h_{\mu\nu}=\eta_{\mu\nu}$, $k=0$ and $\lambda^2= -6\Lambda$.
In the generic case, however, $\Lambda_{4}$ is canceled by the matter contributions encoded 
in the energy-momentum tensor and the dilaton-induced terms on the right-hand side of (\ref{eef}). 
This counterbalance of the four-dimensional vacuum energy by matter cosmological
terms is imposed by the junction conditions representing a different realization of
the self-tuning mechanism of \cite{kach,kachh}. 

Note however, that in the case at hand we are not making any {\it a priori} assumption on the
form of the energy-momentum tensor on the brane $\tau_{\mu\nu}$. Israel junction
condition, together with the jump condition for the dilaton across $\chi=0$,
force it to be of vacuum type, the net cosmological term (\ref{cancel}) being zero 
independently of the value of $\rho_{0}$, $b$ and $\xi$. 

For the five-dimensional metrics of the form (\ref{warp}) we can evaluate the tensor $E_{\mu\nu}$ in terms of the 
curvature and the scalar field. The result is
\begin{eqnarray}
E_{\mu\nu}&=&-{1\over 3}\left(R_{\mu\nu}-{1\over 4}g_{\mu\nu}\,R\right)\nonumber \\
&-&{2k^2\over 27}\left[\partial_{\mu}\varphi
\partial_{\nu}\varphi-{1\over 4}(\partial\varphi)^2 g_{\mu\nu}\right] \nonumber \\
&-&{2k\over 9}\left(D_{\mu}D_{\nu}\varphi-{1\over 4}\,D^2\varphi\,g_{\mu\nu}\right)\, .
\label{emunu}
\end{eqnarray}
Using this expression the effective energy-momentum tensor driving the four-dimensional geometry can be written 
solely in terms of the four-dimensional projection of the scalar field
$\varphi(x)$. The four-dimensional equations 
are given by
\begin{eqnarray}
R_{\mu\nu}&-&\half g_{\mu\nu}R =\partial_{\mu}\varphi\partial_{\nu}\varphi-\half g_{\mu\nu}(\partial\varphi)^2 
\nonumber \\
&+&{k\over 3}\left(D_{\mu}D_{\nu}\varphi-{1\over 4}g_{\mu\nu}\,D^2\varphi\right) \nonumber \\
&+&{k^2\over 9}\left[\partial_{\mu}\varphi\partial_{\nu}\varphi-{1\over 4}g_{\mu\nu}(\partial\varphi)^2\right]
\label{eefd}
\end{eqnarray}

The first important thing to notice is that when $k=0$ we recover an ordinary four-dimensional scalar cosmology
with $\varphi(x)=\sqrt{6}\,Q(x)$. Actually,
by looking at the original ansatz for the vacuum five-dimensional metric in eq. (\ref{sp}) we realize that this is exactly the 
four-dimensional geometry whose ``oxidation'' produces (\ref{sp}). This
is yet another way to see our previous conclusion, that any four-dimensional cosmology coupled
to a massless scalar field can be trivially embedded into a five-dimensional cosmology with negative cosmological constant by
the ansatz (\ref{rsun}). In this case $E_{\mu\nu}$ has a purely ``geometrical'' origin, being determined just by the 
four-dimensional curvature, as can be seen from (\ref{emunu}). 

If $k\neq 0$ things get more involved, however, since now the effective energy-momentum tensor 
in (\ref{eefd}) cannot be interpreted as that of a minimally coupled massless scalar field. The nontrivial cosmological
dynamics in the $\chi$-coordinate translates into Brans-Dicke-like terms in the energy-momentum tensor (cf. \cite{eft}). 
It is interesting to notice that in the large-$|k|$ limit even if the dilaton $\varphi(x)$ is switched off, there
is a residual contribution coming from the last two terms in (\ref{eefd}). Since $\varphi(x)\sim 1/|k|$ for large $|k|$
the weakness of the scalar field is compensated by the extra powers of $k$ in (\ref{eefd}) so a source for the gravitational
field is left in that limit, given by the terms in the ``effective'' energy-momentum tensor with explicit $k$-dependence.          

As it turns out the four-dimensional cosmologies can be conformally related to the
equations of low energy string cosmology. In order to see this 
note that the vacuum Einstein equations for (\ref{sp}) imply that
$D^{2}_{h}\varphi=0$, where the covariant derivative is defined 
with respect to the metric $h_{\mu\nu}$; when written 
in terms of the four-dimensional metric $g_{\mu\nu}=e^{-{k\over 3}\varphi}h_{\mu\nu}$
the condition translates into
$D^2\varphi=-{k\over3}g^{\mu\nu} \partial_{\mu}\varphi\partial_{\nu}\varphi.$ 
In this way, $E_{\mu\nu}$ is given by
$$
E_{\mu\nu}=-\left({1\over3}+{k^2\over9}\right)
\partial_{\mu}\varphi\partial_{\nu}\varphi+
{1\over12}g_{\mu\nu}
(\partial\varphi)^2
-{k\over3}\,D_{\mu}D_{\nu}\varphi,$$
whereas the four-dimensional Einstein equations can be written as
$$R_{\mu\nu}=\partial_{\mu}\varphi\partial_{\nu}\varphi+{k\over3}
D_{\mu}D_{\nu}\varphi+{k^2\over9}\partial_{\mu}
\varphi\partial_{\nu}\varphi.
$$
Dilaton gravity is then recovered through the field redefinitions (cf. \cite{jim}) 
\begin{eqnarray}
\bar{g}_{\mu\nu}&=&\exp\left[{\left({k\over 3}\pm\sqrt{{k^2+6\over 3}}\right)}\varphi\right]
\,g_{\mu\nu} \nonumber \\
\Phi &=& \pm \sqrt{{k^2+6\over 3}}\,\varphi
\end{eqnarray}
so we arrive to the usual equations for dilaton gravity in the string frame (with the normalization 
$\CL_{\rm grav}\sim e^{-\Phi}\bar{R}$)
\begin{eqnarray}
\bar{R}_{\mu\nu}&=&-\bar{D}_{\mu}\bar{D}_{\nu}\Phi\, , \nonumber \\
\bar{D}^2\Phi &=& \bar{g}^{\mu\nu}\partial_{\mu}\Phi\partial_{\nu}\Phi\, .
\label{esf}
\end{eqnarray}

An important stage in the evolution of the early universe is 
inflation. 
As shown above the solutions on the brane can be conformally related
to low energy string cosmology in the string frame. With one
further conformal transformation they, of course, can be
related to the usual Einstein frame in which the metric is just $h_{\mu\nu}$ and
the physics is described by Einstein relativity coupled to a massless
scalar field $\psi(x)_{\rm 4D}=\sqrt{6}\,Q(x)$. In this frame standard inflation
does not take place. However, in the string frame there is
the possibility of implementing inflation in the framework
of the pre-big-bang scenario \cite{gab}. In particular, if the four-dimensional
metric has a space-like Killing vector, scale factor duality along that direction
will be a symmetry of the equations (\ref{esf}). 

It is interesting to note that from the point of view of a
five-dimensional observer there are inflating solutions on the
brane. Using the form of the energy momentum tensor in 5 dimensions
the strong energy condition reads, for any timelike vector $t^{A}$ ($t_{A}t^{A}=-1$),
$$
0<\CR_{AB}t^At^B=(t^A\partial_{A}\phi)^2-{2\over3}\Lambda\, e^{-{2\over 3}k\phi}\, .
$$
This condition can in principle be violated for $\Lambda>0$, i.e.
if $k^2>12$.

\section{Examples}

\subsection{Bianchi I brane cosmologies}

As an illustrating example we can construct Bianchi I brane cosmologies generalizing the 
analysis of \cite{fr} to nonzero profiles for a scalar field with a nontrivial potential. 
We start with a vacuum Kasner-like line element in five dimensions and apply the algorithm described 
to find, after a suitable redefinition of the constants, the following solution (we will
consider that $k^2\neq 3$. The case $k^2=3$ can be easily obtained)
\begin{eqnarray}
ds^2&=&t^{{2k\over 3}\beta} e^{{2k^2\,\xi\over k^2-3}|\chi|}d\chi^2 \nonumber \\
&+&e^{{6\xi\over k^2-3}|\chi|}\left(-dt^2+t^{2\alpha_{1}}dx^2+t^{2\alpha_{2}}dy^2+t^{2\alpha_{3}}dz^2
\right)\nonumber \\
\phi(t,\chi)&=& \beta\log{t}+{3k\xi\over k^2-3}|\chi| \nonumber 
\end{eqnarray}
where the constants $\alpha_{i}$, $\beta$ satisfy the following relations
\begin{equation}
\sum_{i=1}^{3}\alpha_{i}=1-{k\beta\over 3}, \hskip 1cm \sum_{i=1}^{3}\alpha_{i}^{2}=1-\beta^2
-\left({k\beta\over 3}\right)^2 .
\label{constr}
\end{equation}

We notice that for $k=0$ the usual condition for a Kasner metric coupled to a massless scalar field are retrieved.
As we saw in the general analysis of Section 4, for vanishing $k$ the four-dimensional effective energy-momentum tensor
is that of a massless scalar field. When $k\neq 0$ the usual Kasner conditions get modified. Looking at the scaling
of powers of $k\beta$ in (\ref{constr}) we see that the first condition gets modified by the linear terms in the
dilaton field in (\ref{eefd}) whereas the modification of the second condition comes from the quadratic ones.
In the Ho\v{r}ava-Witten case the conditions on the Kasner exponents reduce to the ones found in 
\cite{dabr}.

Looking at the constraints (\ref{constr}) it can be easily seen that there is no volume inflation
for any value of the parameter $k$ and $\beta$, since the average scale factor scales
as $t^{n}$ with $0<n<{2\over 3}$.

The case when $|k|\rightarrow \infty$ is of some interest. In this limit, regularity of the
five-dimensional metric requires that $\beta k\sim {\rm constant}$ so the homogeneous
part of the scalar field vanishes. The result is a Bianchi I metric in four dimensions where
the Kasner exponents $\alpha_{i}$ satisfy the constraints
$$
\sum_{i=1}^3 \alpha_{i}=1-C, \hskip 1cm \sum_{i=1}^3\alpha_{i}^2=1-C^2
$$
with $|C|\leq 1$ a numerical constant. In this case the (traceless) 
effective energy-momentum tensor is that of an anisotropic fluid with 
energy density and pressure given by
$$
\rho={C(1-C)\over t^2}, \hskip 1cm p_{i}={C\alpha_{i}\over t^2}.
$$

\subsection{Friedmann-Robertson-Walker models}

Another physically interesting example are Friedmann-Robertson-Walker (FRW) cosmologies. These can be easily constructed by 
considering the five-dimensional vacuum line element \cite{fvm}
\begin{eqnarray}
ds^2_{\rm vac}&=&{t^2\over 1-\kappa t^2}d\chi^2 \nonumber \\
&-&dt^2+(1-\kappa t^2)\left[{dr^2\over 1-\kappa r^2}+r^2\left(d\theta^2+\sin^2\theta d\phi^2\right)\right]
\nonumber
\end{eqnarray}
where $\kappa=0,\pm 1$. Taking $\chi$ as the bulk coordinate and applying the algorithm of Section 3 we get,
after changing into conformal time, the 
following four-dimensional metrics
$$
ds^{2}_{\rm 4D}=a^{2}(\eta)\left[
-d\eta^2+{dr^2\over 1-\kappa r^2}+r^2\left(d\theta^2+\sin^2\theta d\phi^2\right)\right]
$$
where
\begin{equation}
a(\eta)=\left\{
\matrix{(\cosh{\eta})^{\sqrt{k^2+6}+k\over 2\sqrt{k^2+6}}(\sinh{\eta})^{\sqrt{k^2+6}-k\over 2\sqrt{k^2+6}} & & \kappa=-1 \cr
 & & \cr
\hskip -3.1cm\eta^{\sqrt{k^2+6}-k\over 2\sqrt{k^2+6}} & & \kappa=0 \cr
 & & \cr
\hskip -0.3cm (\cos{\eta})^{\sqrt{k^2+6}+k\over 2\sqrt{k^2+6}}(\sin{\eta})^{\sqrt{k^2+6}-k\over 2\sqrt{k^2+6}} & & \kappa=1}
\right.
\label{sf}
\end{equation}
while the dilaton field is
$$
\varphi(\eta)=\left\{
\matrix{{3\over \sqrt{k^2+6}}\,\log{\tanh{\eta}} & & \kappa=-1 \cr
 & & \cr
\hskip -0.8cm { 3\over \sqrt{k^2+6}}\,\log{{\eta}} & & \kappa=0 \cr
 & & \cr
\hskip -.1cm {3\over \sqrt{k^2+6}}\,\log{\tan{\eta}} & & \kappa=1} 
\right.
$$

In the flat ($\kappa=0$) case we recover the solution discussed in \cite{chsa}. 
There is no inflation for any value of 
$k$, since the exponent of the scale factor in (\ref{sf}) is always positive. It is interesting 
to notice, however, that when $k\rightarrow -\infty$ we recover a radiation-dominated universe. Looking at the 
effective energy-momentum tensor on the brane, eq. (\ref{eefd}), we see that this corresponds to the dynamics of the universe
being dominated by $E_{\mu\nu}$. The opposite limit $k\rightarrow \infty$ gives Minkowski space-time as a result.

For the non-flat FRW models we find again that inflation does not occur for any value of $k$. In the case of 
the negatively curved model ($\kappa=-1$) the universe approaches a radiation dominated regime at late times ($\eta\rightarrow \infty$)
in which the dilaton field is frozen, independently of $k$. As in the flat case, the limit $k\rightarrow-\infty$ retrieves
a radiation dominated open FRW model. When $k\rightarrow \infty$, on the other hand, we get a {\it regular} vacuum FRW model with
scale factor $a(\eta)=\cosh{\eta}$. The conclusions are similar in the case of the models with $\kappa=1$: no value of $k$ renders
an inflationary universe.

We will not  present  further explicit  examples, but just mention again that the ansatz 
(\ref{sp}) is quite generic. One may start directly with any vacuum five-dimensional solution and put
it into (\ref{sp}) form, or rather start with a four-dimensional dilaton solution lifting it to a 
five-dimensional 
vacuum geometry. Thus, for example, the solutions of \cite{hlz}  and their generalizations may be
easily obtained by using the dilatonic plane wave solutions given in \cite{fkvm}.

\section{Conclusions and outlook}

Brane cosmology is special, as compared to standard Einstein gravity, in that the four-dimensional world is not
dynamically self-contained, in the sense that the matter/energy content of the universe encoded in the 
energy-momentum 
tensor does not determine the gravitational field. The ``missing'' part on the right-hand side of the Einstein equations
comes from gravitational effects in the bulk which are not sourced by four-dimensional matter. It is important to notice, 
however, that the term containing the tidal bulk effects is the only one not suppressed by powers of the five-dimensional 
Newton's constant.  

Here we have studied the physical and cosmological relevance of bulk effects by looking at 
five-dimensional 
cosmologies coupled to a Liouville scalar field and embedding the brane world into them. 
One of the remarkable properties of the class of brane cosmologies under study is that
there is a natural self-tuning of the vacuum energy on the brane. This insensitivity of the brane solutions
to the value of the brane tension happens because the 
``bare" cosmological constant on the brane is dynamically counterbalanced by the brane matter, which
the junction conditions force to be of vacuum energy type. So,
the dynamics of the four-dimensional
universe is driven just by the dilaton field and the non-local bulk effects contained in 
the tensor $E_{\mu\nu}$. 

The five-dimensional bulk cosmologies we constructed using the theorem stated in 
Section 3 generalize, and include, those studied previously in the literature, (see, for example, 
\cite{jim,kach,gjs,ajs,hlz,chsa}).
In addition, the brane cosmologies obtained by warped embeddings extend and complement
the self-tuning mechanism of \cite{kach,nem} to generic non Ricci-flat branes.
Incidentally, the technique proposed can be also used to construct static five-dimensional solutions
which on the brane reduce to four-dimensional black holes.

The authors of \cite{flln,fllnd} have argued that the self-tuning mechanism proposed in \cite{kach} is 
actually a fine tuning in disguise due to the presence of the singularity in the bulk. For the brane worlds
studied in this paper, however, we find that whenever $k^2<3$ the four-dimensional effective theory is well 
defined, in the sense that the four-dimensional Planck scale is finite and the consistency condition 
of ref. \cite{fllnd} is satisfied without adding extra sources due to the ``on-shell'' identity
\begin{eqnarray}
{2\over 3}\Lambda \int_{-\infty}^{\infty}d\chi & & \hspace*{-0.5cm}\sqrt{-\CG}\,e^{-{2\over 3}k\phi(\chi,x)}
\nonumber \\
&+&{1\over 3}\sqrt{-g}\left[\lambda(\varphi)+\rho_{0}e^{4b\varphi(x)}\right]=0,
\label{cc2}
\end{eqnarray}
where we have used the fact that for our family of solutions $L_{\rm matter}=-\rho_{0}$. Actually
eq. (\ref{cc2}) is automatically enforced by Israel junction conditions and the matching condition for the dilaton, 
which in turn determine both the energy-momentum tensor on the brane and $\lambda(\varphi)$.
If $k^2\geq 3$ the divergence in the four-dimensional Planck scale can be cut off,
provided a second ``hidden'' vacuum brane 
is located at some $\chi_{0}>0$. The coordinate $\chi$ is then restricted to the interval $[0,\chi_{0}]$
by assuming ${\bf Z}_{2}$ reflection symmetry around the location of the second brane. Imposing the junction 
conditions at $\chi_{0}$ we find $\lambda(\phi)_{\rm hidden}=-\lambda(\phi)$, where 
$\lambda(\phi)$ is given by eq. (\ref{lamb}) with $\rho_{0}=0$. This value of the 
vacuum energy for the second brane implies again that the model satisfies the consistency condition of \cite{fllnd}
for any value of $k$.

In order to illustrate the physics of the family of brane worlds considered, we have analyzed a 
number of explicit examples of physical relevance. For Bianchi-I and all the FRW models we find
that the time-dependence of the scale factor(s) is only controlled by the constant $k$ measuring the 
slope of the potential of the scalar field in the bulk. The result is that no inflation occurs 
on the brane for any value of $k$, whereas there is a region in parameter space for which
inflation takes place in the bulk. The difference between the behavior on the brane and in the
bulk is mainly due to difference in the character of the velocity of the fluid flow as 
derived from the scalar field on the brane and in the bulk. While the
five-velocity in the bulk is not orthogonal to the hypersurfaces of constant time 
its projection onto the brane is.

In the analysis presented here we have assumed that the brane world is trapped at an orbifold point and it does not 
move in the bulk. It can be easily seen that this is a consistent assumption for the self-tuning brane cosmologies
studied in Section 4. Once ${\bf Z}_{2}$ symmetry is relaxed in general the brane will move and
the problem of the dynamical stability of the brane world arises. A specially interesting way
to relax reflection symmetry in our case is to consider bulk space-times in which $\xi$ takes different
values at the two sides of the brane. If we regard the dilaton potential as arising from some 
kind of phase transition, the brane world plays the role of a (infinitely thin) domain wall separating 
two regions of space with different values for the cosmological constant, in the spirit 
of the scenarios discussed in \cite{bu}. In this case the possible motion of the brane in the bulk 
will affect the gravitational dynamics on the brane world \cite{krauss,stw}. A detailed study
of these non-symmetric brane worlds and their stability will be presented elsewhere.

\begin{acknowledgments}

One of us (M.A.V.-M.) would like to thank Jun Nishimura for stimulating discussions.  
A.F. acknowledges support from University of the Basque Country Grants UPV 172.310-EB150/98 and
UPV 172.310-G02/99 and Spanish Science Ministry Grant 1/CICYT 00172.310-0018-12205/2000. K.E.K. 
has been supported by the Swiss National Science Foundation and Spanish Science Ministry Grant 
1/CICYT 00172.310-0018-12205/2000. The work of M.A.V.-M. has been
supported by EU Network ``Discrete Random Geometry'' Grant HPRN-CT-1999-00161, ESF Network no. 82
on ``Geometry and Disorder'', Spanish Science Ministry Grant AEN99-0315 and 
University of the Basque Country Grants UPV 063.310-EB187/98 and UPV 172.310-G02/99.

\end{acknowledgments}


\begin{thebibliography}{99}

\bibitem{rubshap}
K. Akama, {\it Pregeometry}, in: ``Gauge Theory and Gravitation'', eds. 
K. Kikkawa, N. Nakanishi and H. Nariai, Springer-Verlag 1983 (\bb{0001113});\\
V. Rubakov and M. Shaposhnikov, \plb{125}{1983}{136.}

\bibitem{hw}
P. Ho\v{r}ava and E. Witten, \npb{460}{1995}{506} (\bb{9510209}).

\bibitem{lardim}
N. Arkani-Hamed, S. Dimopoulos and G. Dvali, \plb{429}{1998}{263} (\bb{9803315});\\
I. Antoniadis, N. Arkani-Hamed, S. Dimopoulos and G. Dvali, \plb{436}{1998}{257} (\bb{9804398}).


\bibitem{rs}
L. Randall and R. Sundrum, \prl{83}{1999}{3370} (\bb{9905221});
\prl{83}{1999}{4690} (\bb{9906064})

\bibitem{gt}
J. Garriga and T. Tanaka, \prl{84}{2000}{2778} (\bb{9911055}).

\bibitem{phen}
R. Sundrum, \prd{59}{1999}{085009} (\heph{9805471});\\
N. Arkani-Hamed, S. Dimopoulos and G. Dvali, \prd{59}{1999}{086004} (\heph{9807344});\\
H. Davoudias, J.L. Hewett and T.G. Rizzo, \prl{84}{2000}{2080} (\heph{9909255});\\
I.I. Kogan, S. Mouslopoulos, A. Papazoglou, G.G. Ross and J. Santiago, \npb{584}{2000}{313} (\heph{9912552});\\
S.B. Bae and H.S. Lee, \plb{487}{2000}{299} (\heph{0002224});\\
A.-C. Davis, S.C. Davis, W.B. Perkins and I.R. Vernon, \plb{504}{2001}{254} (\heph{0008132});\\
K. Cheung, \prd{63}{2001}{056007} (\heph{0009232}).


\bibitem{mwbh}
R. Maartens, D. Wands, B.A. Basset and I. Heard, \prd{62}{2000}{041301}
(\heph{9912464}).


\bibitem{cll}
E.J. Copeland, A.R. Liddle and J.E. Lidsey, \prd{64}{2001}{023509} (\aph{0006421}).

\bibitem{bc}
A. Lukas, B.A. Ovrut and D. Waldram, \prd{60}{1999}{086001} (\bb{9806022});
\prd{61}{2000}{023506} (\bb{9902071});\\
G. Dvali and S.H. Henry Tye, \plb{450}{1999}{72} (\heph{9812483});\\
P. Binetruy, C. Deffayet and D. Langlois, \npb{565}{2000}{269} (\bb{9905012});\\
E.E. Flanagan, S.H. Henry Tye and I. Wasserman, \prd{62}{2000}{024011} (\bb{9909373});\\
A. Kehagias and E. Kiritsis, \jhep{11}{1999}{022} (\bb{9910174});\\
D. Ida, \jhep{09}{2000}{014} (\grqc{9912002});\\
C. Barcel\'o and M. Visser, \plb{482}{2000}{183} (\bb{0004056});\\
S. Nojiri, O. Obreg\'on and S.D. Odintsov, \prd{62}{2000}{104003} (\bb{0005127};\\
D. Youm, \mpla{16}{2001}{937} (\bb{0012246});\\
P. Binetruy, C. Deffayet and D. Langlois, {\it The radion in brane cosmology}, \bb{0101234};\\
A. Hebecker and J. March-Russell, {\it Randall-Sundrum II cosmology, AdS/CFT, and the bulk black
hole}, \heph{0103214};\\
J. Khoury, B.A. Ovrut, P. Steinhardt and N. Turok, {\it The ekpyrotic universe: colliding branes and 
the origin of the hot big-bang}, \bb{0103239};\\
R. Kallosh, L. Kofman and A. Linde, {\it Pyrotechnic universe}, \bb{0104073};\\
A. Kehagias and K. Tamvakis, {\it A note on brane cosmology}, \heph{0104195}.

\bibitem{hl}
H.A. Chamblin and H.S. Reall, \npb{562}{1999}{133} (\bb{9903225});\\
R.M. Hawkins and J.E. Lidsey \prd{63}{2001}{041301} (\grqc{0011060}).

\bibitem{mss}
R. Maartens, V. Sahni and T.D. Saini, \prd{63}{2001}{063509} (\grqc{0011105}).

\bibitem{svf}
M.G. Santos, F. Vernizzi and P.G. Ferreira, {\it Isotropization and instability
of the brane}, \heph{0103112}.

\bibitem{hlz}
G.T. Horowitz, I. Low and A. Zee, \prd{62}{2000}{086005} (\bb{0004206}).

\bibitem{fr}
A.V. Frolov, {\it Kasner-AdS spacetime and anisotropic brane-world cosmology,}
\grqc{0102064}.

\bibitem{csop}
A. Campos and C.F. Sopuerta, {\it Bulk effects in the cosmological 
dynamics of brane-world scenarios}, \bb{0105100.}

\bibitem{wein}
S. Weinberg, \rmp{61}{1989}{1;} {\it The cosmological constant problems,} talk at
Dark Matter 2000, \heph{0005265}.

\bibitem{strau}
N. Straumann, {\it On the mistery of the cosmic vacuum energy density,}
talk at DARK 2000, \aph{0009386}.

\bibitem{kach}
S. Kachru, M. Schulz and E. Silverstein, \prd{62}{2000}{045021} (\bb{0001206}).

\bibitem{kachh}
S. Kachru, M. Schulz and E. Silverstein, \prd{62}{2000}{085003} (\bb{0002121}).

\bibitem{nem}
N. Arkani-Hamed, S. Dimopoulos, N. Kaloper and R. Sundrum,
\plb{480}{2000}{193} (\bb{0001197}).

\bibitem{cm}
S.M. Carroll and L. Mersini, {\it Can we live in a selftuning universe?}, 
\bb{0105007}.

\bibitem{mw}
K. Maeda and D. Wands, \prd{62}{2000}{124009} (\bb{0008188}).


\bibitem{wald}R.M. Wald, {\it General Relativity}, Chicago 1984.

\bibitem{sms}
T. Shiromizu, K. Maeda and M. Sasaki, \prd{62}{2000}{024012} (\grqc{9910076}).

\bibitem{mb}A. Mennim and R.A. Battye, \cqg{18}{2001}{2171} (\bb{0008192}).


\bibitem{roy}R. Maartens, {\it Geometry and dynamics of the brane world}, talk at the 2000 Spanish 
Relativity Meeting, \grqc{0101059}.

\bibitem{bcmu}
R.A. Battye, B. Carter, A. Mennim and J.-P. Uzan, {\it Einstein equations for an asymmetric
brane world}, \bb{0105091}.

\bibitem{is}
W. Israel, \nc{B44}{1966}{1.}

\bibitem{fonarev}
O. Fonarev, \cqg{12}{1995}{1739} (\grqc{9409020}).

\bibitem{jim}
J.E. Lidsey, \cqg{17}{2000}{L39} (\grqc{9911066}).

\bibitem{ajs}
N. Alonso-Alberca, B. Janssen and P.J. Silva, \cqg{17}{2000}{L163}
(\bb{0005116}).

\bibitem{losw}
A. Lukas, B.A. Ovrut, K.S. Stelle and D. Waldram, \prd{59}{1999}{086001} 
(\bb{9803235}).

\bibitem{fvm}
A. Feinstein and M.A. V\'azquez-Mozo, \npb{568}{2000}{405} (\bb{9906006}).


\bibitem{rh}
A. Chamblin, S.W. Hawking and H.S. Reall, \prd{61}{2000}{065007} (\bb{9909205}).

\bibitem{chsa}H. Ochiai and K. Sato, \plb{503}{2001}{404} (\bb{0010163}).

\bibitem{eft}
D. Youm, \prd{63}{2001}{047503} (\bb{0001166});\\
Y.S. Myung and H.W. Lee, \prd{63}{2001}{064034} (\bb{0001211});\\
C. Barcel\'o and M. Visser, \jhep{10}{2000}{019} (\bb{0009032}).

\bibitem{gab}
G. Veneziano, {\it String cosmology: The pre-big bang scenario}, in: ``The Primordial
Universe", proceedings to the 1999 Les Houches Summer School, eds. P. Binetruy, R. Schaeffer, 
J. Silk and F. David. Springer-Verlag 2001 (\bb{0002094}).


\bibitem{dabr}
M.P. D\c{a}browski, \plb{474}{2000}{226} (\bb{9911217}).

\bibitem{fkvm}
A. Feinstein, K.E. Kunze and M.A. V\'azquez-Mozo, \cqg{17}{2000}{3599} (\bb{0002070}).

\bibitem{gjs}
C. G\'omez, B. Janssen and P. Silva, \jhep{04}{2000}{024} (\bb{0002042}).

\bibitem{flln}
S. F\"orste, Z. Lalak, S. Lavignac and H.P. Nilles, \plb{481}{2000}{360} (\bb{0002164}).

\bibitem{fllnd}
S. F\"orste, Z. Lalak, S. Lavignac and H.P. Nilles, \jhep{09}{2000}{034} (\bb{0006139}).


\bibitem{bu}
B. Carter and J.-P. Uzan, {\it Reflection symmetry breaking scenarios with minimal gauge 
form coupling in brane world cosmology}, \grqc{0101010}.


\bibitem{krauss}
P. Kraus, \jhep{12}{1999}{011} (\bb{9910149}).

\bibitem{stw}H. Stoica, S.-H. Henry Tye and I. Wasserman, \plb{482}{2000}{205} (\bb{0004126}).















\end{thebibliography}
\end{document}